\documentclass[a4paper]{article}

\usepackage{graphics,epsfig}

\newcommand{\gaa}{d}
\newcommand{\gan}{d_{n,F-}}
\newcommand{\f}{D_{2}}
\newcommand{\fn}{D_{k}}
\newcommand{\fnn}{D_{2,n}}
\newcommand{\dd}{s}
\newcommand{\laa}{\lambda_m}
\newcommand{\omm}{\Omega_l}

\newcommand{\omn}{\Omega_{l,n}}

\newcommand{\be}{\begin{equation}}
\newcommand{\ee}{\end{equation}}

\usepackage{amsmath}
\usepackage{amssymb}
\usepackage{amsfonts}

\begin{document}

\title{Chaotic, staggered and polarized dynamics
in opinion forming: the contrarian effect}
\author{Christian Borghesi$^1$\thanks{E-mail:
christian.borghesi@cea.fr}\;\; and\; Serge Galam$^2$\thanks{E-mail:
serge.galam@polytechnique.edu}\\
$^1$Service de Physique de l'Etat CondensŽ (CNRS URA 2464), \\CEA Saclay,
91191 Gif sur Yvette Cedex, France\\
$^2$Centre de Recherche en \'Epist\'emologie Appliqu\'ee,\\
CREA-\'Ecole Polytechnique (CNRS UMR 7656),\\
1, rue Descartes, F-75005 Paris, France}
\date{(15 Mar 2006)}
\maketitle

\begin{abstract}

We revisit the no tie breaking 2-state Galam contrarian model of
opinion dynamics for update groups of size 3. While the initial
model assumes a constant density of contrarians $a$ for both
opinions, it now depends for each opinion on its global support.
Proportionate contrarians are thus found to indeed preserve the
former case main results. However, restricting the contrarian
behavior to only the current collective majority, makes the
dynamics more complex with novel features. For a density
$a<a_c=1/9$ of one-sided contrarians, a chaotic basin is found in
the fifty-fifty region separated from two majority-minority point
attractors, one on each side. For $ 1/9<a\lesssim 0.301 $ only the
chaotic basin survives. In the range $a>0.301$ the chaotic basin
disappears and the majority starts to alternate between the two
opinions with a staggered flow towards two point attractors. We
then study the effect of both, decoupling the local update time
sequence from the contrarian behavior activation, and a smoothing
of the majority rule. A status quo driven bias for contrarian
activation is also considered. Introduction of unsettled agents
driven in the debate on a contrarian basis is shown to only shrink
the chaotic basin. The model may shed light to recent apparent contradictory elections with on the one hand very tied results like in US in 2000 and in Germany in 2002 and 2005, and on the other hand, a huge majority like in France in 2002.

\end{abstract}



\section{Introduction}

In the last years the study of opinion dynamics has attracted a
growing number of works
\cite{italy,huber,espagnol,herrmann,stauffer-m,slanina,rumor,redner,mino,frank,weron,chopard} making it a major current trend
of sociophysics \cite{strike, testi}. Most models consider 2-state
opinion agents combined with some local opinion update rule which
implements the dynamics. They are found to lead to an opinion
polarization of the whole population along one of the two
competing opinions. A unifying frame was proposed to incorporate
all these models \cite{uni}. Continuous opinion models yield
similar tendency \cite{continu-w,continu-h}.

More recently, the concept of contrarian was introduced to account
for some peculiar behavior of agents \cite{contra}. A contrarian
is undistinguishable to others, i.e. its opinion evolves also by
local rule updates. However, once it leaves the update group, it
changes individually its opinion to the other one. The shift is
independent of the opinion itself. A contrarian is not a permanent
individual state. Each agent has a probability $a$ to behave like
a contrarian and $(1-a)$ to stick to its opinion. After each cycle
of local updates, on average a proportion $a$ of agents shifts
spontaneously their opinion to the other one, while a proportion
$(1-a)$ sticks to its current opinion. Given a fixed density of
contrarians $a$, the associated opinion dynamics is then studied
\cite{contra}.

As intuitively expected, the existence of contrarians was shown
to avoid total opinion polarization with the creation of stable
attractors characterized by a stable coexistence between a large
majority and a small minority. However, above some low density,
they were found to produce an unexpected reversal of the dynamics,
with the merger of the two attractors at the former separator,
turning it to the unique stable attractor \cite{contra,contra-s}.
Accordingly, for whatever initial conditions the dynamics leads to
an exact global balance between the two competing opinions. It
thus offers a possible explanation to recently observed hung
election scenario \cite{contra}. A hung election being a two candidate run for which the result is very tied around fifty percent. Chaotic regime was also found in
the description of investors in stock markets \cite{corcos}.

But in today campaigns, polls are regularly publicized making
agents aware of which opinion is currently leading at the global
level. It is therefore more natural to link the propensity to a
contrarian behavior to the current level of global support for a
given opinion. Accordingly, in this paper we relax above
constraints of fixed independent contrarian density $a$ in two
successive steps. First, we study the effect on the dynamics
making the density $a$ proportional for each opinion to its
current global support. Contrarians become proportionate
contrarians. Second, we push the asymmetry further by restricting
the contrarian behavior to only the current majority opinion,
contrarians being one-sided. While proportionate contrarians are
found to preserve the mean features of the fixed density
contrarian dynamics, a more complex situation including a chaotic
regime is discovered for one-sided contrarians.

For a density $a<a_c=1/9$, one-sided contrarians produce a chaotic
basin located in the fifty-fifty percent region. The associated
Lyapunov exponent is calculated. However, there still exist
majority-minority coexistence
point attractors located on each side of the chaotic basin.
Initial conditions determines which regime will dominate, either chaotic outcome around fifty percent,
or point attractor with a well defined majority.
For $1/9<a\lesssim 0.301 $ only the chaotic basin survives. Further, in the range $a>0.301$.
the chaotic basin disappears and the majority starts to alternate
between the two opinions with a staggered flow towards two
majority-minority point attractors.

On this basis the effect of additional social parameters are
studied. A constant shift for contrarian activation is
considered. Decoupling the local update time sequence from the
contrarian behavior, and a delay in accounting for a change of
majority side are also introduced. Last, unsettled
agents are included. They are found to only shrink the chaotic basin making the outcome very tied.

Given above framework, the results
may shed light to recent very unusual hung elections like in the 2000
US presidential vote and in Germany for the 2005 elections. In both cases the outcomes were very tied with almost identical support. At the same time, the model can also provides
an explanation for the 2002 French
presidential election where the winner obtained a huge majority around $80\%$.

At this stage it is worth to stress that we are able to embody these contradictory voting outcomes within a single frame. But at the same time, we are not able to decide before hand which one will prevail. To overpass this difficulty would require the collaboration with social scientists to estimate the actual type of contrarians involved in a given election.

The rest of the paper is organized as follows. In the next Section
we review the original Galam model of contrarians. Throughout the
paper, the size of update local groups is kept equal to three
agents. The contrarians are then made proportionate in Section 3.
Section 4 considers one-sided contrarians and the corresponding
complex dynamics topology. Novel features are obtained.
In Section 5 we study the effect of
decoupling the local update time sequence from the contrarian
behavior activation. A smoothing of the majority rule is also
included. The effect of a status quo driven bias for contrarian
activation is investigated in Section 6.
Section 7 deals with the case of unsettled agents.
Last section contains some discussion.


\section{The original Galam model: individual contrarians }

We start recalling the Galam model of 2-state opinion dynamics
model extended to the presence of contrarians \cite{contra}. It
considers a population of $N$ agents where at a time $t$, $N_A(t)$
persons support one opinion $A$ and $\{N-N_A(t)\}$ persons support
another competing opinion B. In terms of global proportions among
the whole population, it yields $p_{t}=\frac{N_A(t)}{N}$ for $A$
and $\{1-p_{t}\}$ for $B$. These values can be evaluated at any
time using polls.

From an initial value $p_{t}$ at time $t$, a dynamics is
implemented in two steps. First, a neighborhood step where agents
are distributed randomly among various size groups in which they
update their respective individual opinion following the local
initial majority \cite{mino}. In case of a tie in even size
groups, either one opinion is adopted according to some
probabilities \cite{hetero}. The step is accounted by a discrete
time increment of $+1$ leading to a new proportion $p_{t+1}$ of
agents supporting opinion $A$. The second step is contrarian, each
agent individually either shifts its respective opinion to the
other one with a probability $a$, or preserves its current opinion
with probability $(1-a)$. This second step yields an additional
increment of time $+1$ and modifies $p_{t+1}$ to another value
$p_{t+2}$.

Throughout this paper we explicit the calculations for the case of
local groups with the unique value 3. Above rules thus yield
respectively for the first step $p_{t}\rightarrow
p_{t+1}=P_m(p_{t})$ with \be\label{e1} p_{t+1}=P_m(p_{t}) \equiv
p_{t}^3+3p_{t}^2\big(1-p_{t}\big), \ee and for the contrarian
second step $p_{t+1} \rightarrow p_{t+2}=P_c(p_{t+1})$ where
\be\label{e2} p_{t+2}=P_c(p_{t+1}) \equiv
(1-a)p_{t+1}+a[1-p_{t+1}]. \ee It should be stressed that before
performing another cycle of opinion updates, agents are reshuffled
\cite{new}. Applying above 2-step cycle repeatedly $n$ times
results in a proportion $p_{t+2n}$ of agents supporting $A$. All
possible variations and extensions of the model can be included
within a unifying frame \cite{uni}.

At this stage we make a change of variable from $p$ to $d$ with $p=d+\frac{1}{2}$. It will appear to
be more convenient for our investigation. A positive $\gaa$
makes $A$ the majority opinion while a negative value grounds it
as minority with a deficit $|\gaa|$ of support with respect to
$B$. In terms of the new variable $\gaa$, $P_m(p_{t})$ and
$P_c(p_{t+1})$ become respectively \be\label{eDm}
\gaa_{t+1}=D_m(\gaa_{t}) \equiv -2\gaa_{t}^3+\frac{3}{2}\gaa_{t},
\ee and \be \gaa_{t+2}=D_c(\gaa_{t+1}) \equiv (1-2a) \gaa_{t+1},
\ee which combine for one full cycle into the single Equation $
\gaa_{t+2}=D_c[D_m(d_{t})]$, which we denote by \be\label{d2}
\gaa_{t+2}=\f(d_{t})=(1-2a)\{-2\gaa_{t}^3+\frac{3}{2}\gaa_{t}\},
\ee where index 2 of $\f$ means one local rule followed by one
contrarian step with a time interval of 2 for the dynamics. At
this stage such a 2-step split could appear artificial but it will
become instrumental latter on to extend the dynamics to cases
where one cycle is built out of $(k-1)$ consecutive local updates
followed by one contrarian effect. These cases will be denoted by
$\gaa_{t+k}=\fn(\gaa_{t})$ with $k$ being the appropriate time
interval to study the associated properties of the dynamics. In
the original Galam work $k=2$.

For this last case the main results obtained from Eqs. (1,2) or
(3,4) are twofold. At low concentration $a$ of contrarian
behavior, total polarization is prevented with two mixed
attractors at which a majority and a small minority coexist. The
threshold for $A$ victory is at $p_v=\frac{1}{2}$ or $\gaa_v=0$
which defines the separator of the dynamics. Furthermore,
increasing $a$ provokes a continuous phase transition at
$a_c=\frac{1}{6}$ turning $p_v=\frac{1}{2}$ or $\gaa_v=0$ into the
unique and stable attractor of the dynamics. The final state is a
perfect equality of both opinions at the collective level with on
going individual opinion shifts \cite{contra}.


\section{Making contrarian behavior opinion current status dependent:
proportionate contrarians}

While the original model considers a constant and fixed proportion
$a$ of contrarians, it seems more realistic to make it depend on
the current state of the system, in particular due to the existence
of published polls.
We thus suppose that if at some specific time $t$ all
agents are informed of the actual value of $d_{t}$, they react
accordingly as contrarians with respect to
$d_{t}$ at time $(t+1)$ leading to $d_{t+1}$. But it thus becomes
natural to make the contrarian behavior opinion dependent. Not to the
opinion itself but to its current level of support in the population.

To distinguish the associated contrarians from previous ones, we
call them proportionate contrarians. Therefore, agents sharing
opinion A react to $p_{t}$, i.e., $d_{t}$ while agents sharing
opinion B react to $(1-p_{t})$, i.e., $-d_{t}$. We denote these
rates respectively $a(d)$ and $b(d)$. From symmetry we have
$b(d)=a(-d)$ since opinions are time reversal.

For the time being, we keep our 2-step cycle, which implies a
regular and periodic publication of polls. Such a constraint will
be relaxed in a latter Section. On this basis the proportionate
contrarian density $a$ becomes a function of time through the
variable $d_{t}$. If no agent shares the opinion A, no one will
react against it as a proportionate contrarian yielding the
constraint $a(d=-1/2)=0$. On the other extreme at $d=1/2$,
$a(d=1/2)=a_0$ where $a_0$ is the proportionate contrarian maximal
value which satisfies $0\leq a_0 \leq 1$.

Keeping in mind that now the proportionate contrarian density is
not the same at a given time among agents sharing respectively
opinion A and B, Eq. (4) becomes \be\label{cont-dens}
\gaa_{t+2}=\f(\gaa_t)=(1-a_{t+1}-b_{t+1})\gaa_{t+1}+
\frac{b_{t+1}-a_{t+1}}{2}, \ee where $a_{t+1}$ and $b_{t+1}$
respectively mean $a(d_{t+1})$ and $b(d_{t+1}$), and $\gaa_{t+1}$
is given by Eq. (3).

From Eq.(6) we can extract several properties of the associated
dynamics. First we note that $d=0$ is a fixed point if and only if
$a(d=0)=b(d=0)$. It means no agent is contrarian at perfect
equality of opinions. Second we can evaluate its stability by
studying the value of the associated eigenvalue $\lambda$ with
respect to one. When $d=0$ is a fixed point, it is an attractor
when $\lambda <1$, which in turn implies the condition \be 6
a(d=0)+a'(d=0)-b'(d=0)>1,\ee where the prime means a derivative
with respect to $d$. Otherwise when $6 a(d=0)+a'(d=0)-b'(d=0)<1$
($\lambda>1$),
the fixed point $d=0$ is a separator. A separator implies the
existence of two attractors located respectively on both side at
$d>0$ and $d<0$ with a stable coexistence of a majority and a
minority.

Using above results we can review few specific functional forms
for the $a$ $d$-dependence. We start with the linear dependence
$a=a_0 p=a_0(1/2+d)$. For $a_0<1/5$, $d_v=0$ is a separator and
the associated attractors are located at $d=\pm
\frac{\sqrt{-1+5a_0}}{2\sqrt{-1+a_0}}$. When $a_0\geq 1/5$, $d=0$
becomes the unique attractor of the dynamics. In other words, the
former Galam result is recovered with proportionate contrarians
stabilizing a perfect equality between both opinions once their
density is larger than a critical value, here $1/5$ \cite{contra}.
Results are shown in Fig. (1).


\begin{figure}
\centerline{ \epsfxsize=7cm\epsfbox{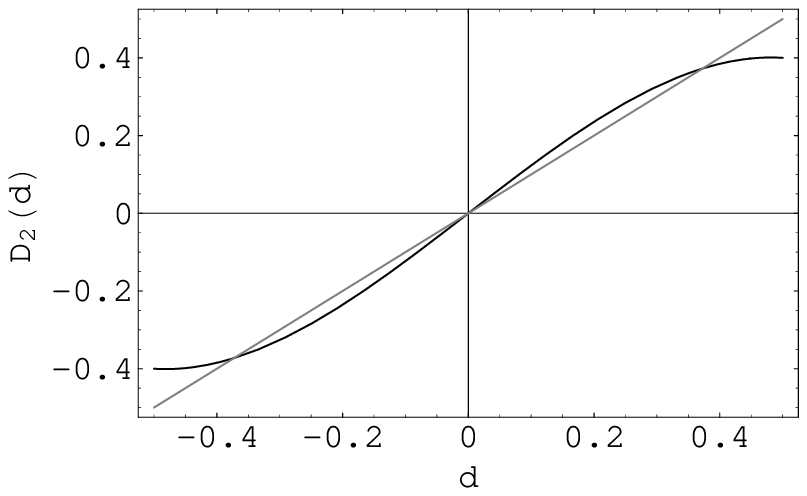}\hspace{1cm}
\epsfxsize=7cm\epsfbox{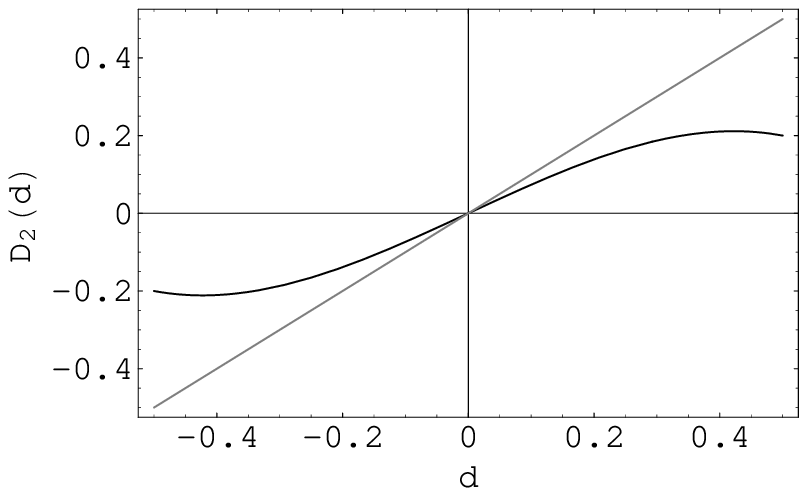}} \caption [ ] {Application $\f$ given by Eq. (\ref{cont-dens}) for the linear and
symmetric case where $a(d)=a_0 p=a_0(1/2+d)$ and $b(d)=a(-d)$. On
the left side $a_0=0.1<1/5$ with two attractors at coexistence of
a majority and a minority. On the right side $a_0=0.3>1/5$ with
one attractor at a perfect balance among both opinions.}
\end{figure}

Considering a power law form $a=a_0 p^\gamma=a_0(1/2+d)^\gamma$ we
get the condition $a_0\geq \frac{2^{\gamma -1}}{3+2\gamma}$ to
make $d=0$ the unique attractor. It is a separator with two mixed
phases attractors when $a_0<\frac{2^{\gamma -1}}{3+2\gamma}$.
Taking $\gamma= 2$ yields a square dependence with $\frac{2}{7}$
for the critical value of $a_0$. More proportionate contrarians
are needed ($a_0\geq \frac{2}{7}=0.286$) with respect to the
linear case ($a_0>\frac{1}{5}=0.20$) to produce the perfect
balance of opinions. When $d=0$ is a separator, the two attractors
are located at $d=\pm \frac{1}{2}
\sqrt{1+\frac{1}{a_0}(1+\sqrt{1+8a_0^2})}$. On the opposite,
taking a square root dependence with $\gamma= 1/2$, yields
$\frac{1}{4\sqrt{2}}\simeq 0.177$ for the critical value of $a_0$
making less proportionate contrarians needed to get the opinion
balance as the unique attractor of the dynamics. The former Galam
model considers $a=cst.$ giving a critical value of
$\frac{1}{6}\simeq 0.167$ which is the lowest value we can get.

At this stage we can conclude that proportionate contrarians do
not modify qualitatively the former constant contrarian density
dynamics opinion.


\section{Restricting contrarian behavior to the current majority:
one-sided contrarians}

On the basis of above results, we go back to the case of a
constant density of contrarians, but now restricting the
activation of the contrarian behavior to only the current majority
opinion. It thus yields for $d>0$ the conditions $a(d>0)=a=cst$
(with $0\leq a\leq 1$) and $b(d>0)=0$, while for $d<0$ the
conditions are $a(d<0)=0$ and $b(d<0)=a=cst$, where we have
assumed symmetric conditions for both opinions. It is worth to
note that this symmetry induces at $d=0$ the condition
$a(d=0)=b(d=0)=\frac{a}{2}$ both preserving the same total density
$a$ of contrarians and recovering here the former original
individual contrarian case. We call these contrarians one-sided
contrarians. The associated rule update writes:
\be\label{eDo-s}\gaa_{t+2}=D_{o-s}(\gaa_{t+1})\equiv(1-a)\gaa_{t+1}-\frac{a}{2}\;sign[\gaa_{t+1}],\ee
where $sign[x]=1$ if $x>0$, $-1$ if $x<0$ and $0$ if $x=0$.

Using Eq. (\ref{eDm}) for $\gaa_{t+1}$ and noting that
$sign[\gaa_{t+1}]=sign[\gaa_t]$, Eq. (\ref{cont-dens}) writes:
\be\label{e5}\gaa_{t+2}=\f(\gaa_t)=(1-a)(\frac{3}{2}\;\gaa_t-2\gaa_t^3)
-\frac{a}{2}\;sign[\gaa_{t}].\ee At second order in $\gaa_t$ it
yields: \be\label{e6} \f(\gaa_t)\simeq\lambda_m
\gaa_t-\frac{a}{2}\;sign[\gaa_t], \ee where
$\laa=\frac{3}{2}(1-a)$ is the maximal slope of the application
$\f$.


\begin{figure}
\centerline{ \epsfxsize=7cm\epsfbox{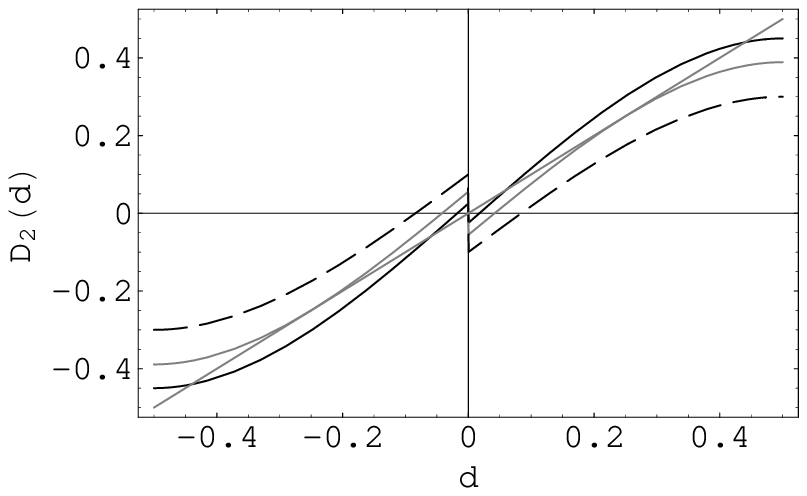}\hspace{1 cm}
\epsfxsize=7cm\epsfbox{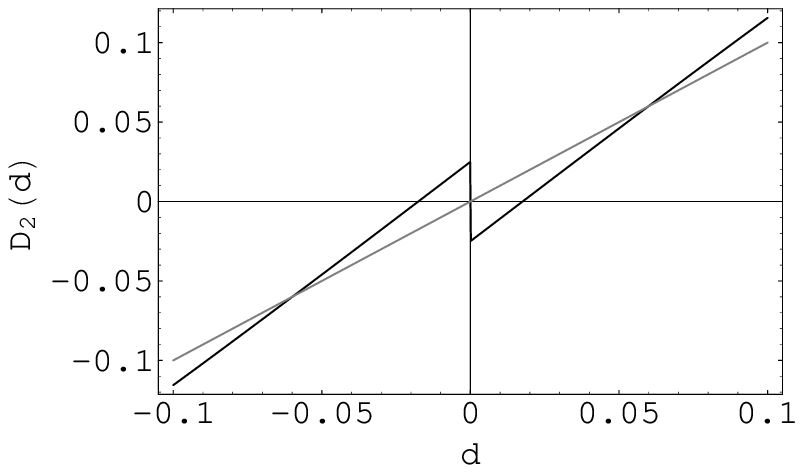}} \caption[ ]{Application $\f$ given by Eq. (\ref{e5}). Left side: for
respectively $a=0.05$ in plain line, $a=a_c=1/9$ in thin line and
$a=0.2$ in dashed line. Right side: for $a=0.05$, a zoom around
$\gaa=0$.}
\end{figure}

We can now study the associated fixed points and their stability.
We first note that the point $d=0$ is a singular fixed point.
Then, the application being symmetric we restrict the study to
$\gaa>0$. According to Eq. (\ref{e5}), fixed points exist if
$a\leq a_c=\frac{1}{9}\simeq 0.11$ and their values are:
\be\label{fixes}\gaa_{F\pm}=\frac{1}{4}(1\pm\sqrt{\frac{1-9a}{1-a}}\;).\ee
When they exist, $\gaa_{F-}$ is unstable and $\gaa_{F+}$ stable
(see Fig. 2). From Eq. (\ref{e6}) it gives:
\be\label{e7}\gaa_{F-}\simeq \frac{a}{2(\laa-1)},\ee with the
eigenvalue $\lambda\simeq\laa>1$ ($\laa>1$ for $a\leq a_c$).

Eq. (\ref{e6}) exhibits clearly condition for expansion when
$\laa>1$, and folding up by the discontinuity at the origin
$\gaa=0$. Thus, for being chaotic, the application $\f$ has to
satisfy two conditions; first, the expansion condition and second,
the perpetuity of the chaotic basin \cite{pomeau,ott}. If
the application $\f$ is chaotic, the interval of successive
iterated points after transients is called $\omm$. Here $\omm
=]-a/2;a/2[$ (see Fig. 2).

$\underline{1^{st}\;condition}$: expansion.

Inside the interval $\omm$ the application $\f$ possesses a slope
$\lambda$, such as $\laa>\lambda>\f'(a/2)$, where $\f'$ denotes
the derivative of $\f$ relating to $\gaa$. This implies that
necessary $\laa>1$, i.e. $a<1/3$. Furthermore, $\f'(a/2)=1$ for
$a\simeq 0.27$. According to this unique condition, $\f$ loses its
chaotic nature for $a$ including to the two
last values.

$\underline{2^{nd}\;condition}$: perpetuity of the chaotic
basin, i.e. $\f\big(\omm\big)\subset \omm$.

If there are not fixed points, this condition is automatically
satisfied. If fixed points exist, i.e. for $a\leq a_c$, they must
be outside the interval $]-a/2;a/2[$, i.e. $\gaa_{F-}>a/2$.
According to Eq. (\ref{e7}) it implies $\laa<2$. This condition is
always satisfied here for group size $3$ while it is not the case
for larger update groups from size $5$. Nevertheless, to observe a
chaotic behavior initial condition must satisfy
$|\gaa(t=0)|<\gaa_{F-}$. Otherwise, the successive iterated points
go to a stable fixed point at $\pm\gaa_{F+}$.

A chaotic behavior is numerically observed until $a\simeq 0.301$.
Illustrations are shown in Fig. (3) for $a=0.2$ and $a=0.05$ (with
an appropriate initial condition in the latter case).

Fig. (4) shows sensitivity to initial conditions and permits the
evaluation of the Lyapunov exponent $\lambda_{lyap}$. The initial
difference $\delta (0)$ grows exponentially, $\delta
(n)\simeq\delta (0)\; e^{n\lambda_{lyap}}$, until saturation at
the typical size of the interval $\omm$. The Lyapunov exponent is
positive with $\lambda_{lyap}\simeq \ln(\laa)$ as seen from Eq.
(\ref{e6}).

The notation adopted in these figures is to consider the $n^{th}$
iteration of the application $\f$ since the beginning of the
electoral campaign as the equivalent time $n$. Thus
$\gaa_{t_0+2n}\equiv \gaa(n)$, where $t_0$ is the time at the
beginning of the electoral campaign. In other words,
$\gaa(n+1)=\f\big[\gaa(n)\big]$.


\begin{figure}
\centerline{ \epsfxsize=7cm\epsfbox{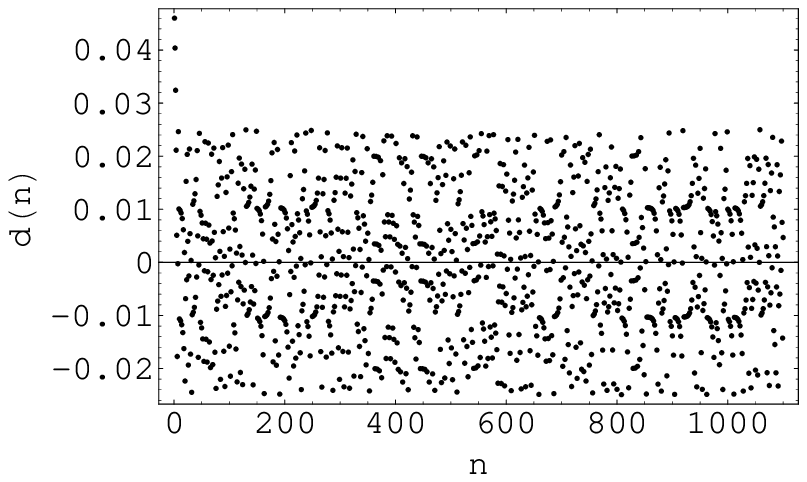}\hspace{1 cm}
\epsfxsize=7cm\epsfbox{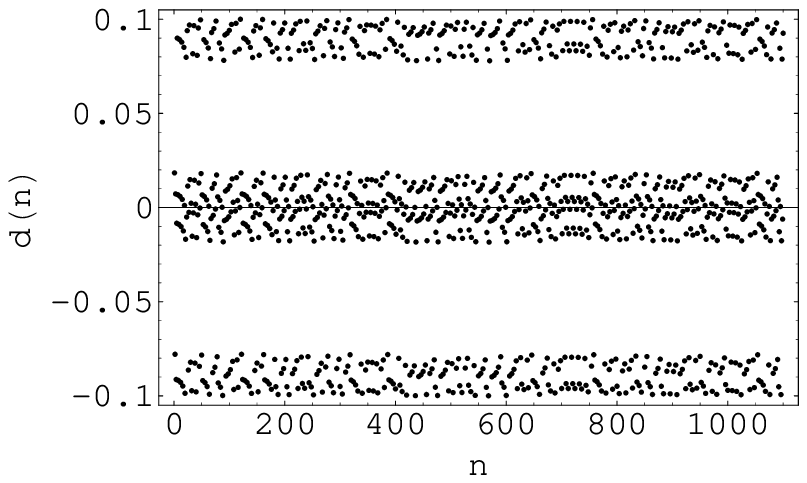}} \caption[ ] {Successive
iterated points by the application $\f$ given by Eq. (\ref{e5}).
Notation here: $\gaa(n+1)=\f\big[\gaa(n)\big]$. Left side:
$a=0.05$ with the initial value $\gaa(0)=0.05$. Right side:
$a=0.2$ with the initial value $\gaa(0)=0.1$.}
\end{figure}


\begin{figure}
\centerline{ \epsfxsize=7cm\epsfbox{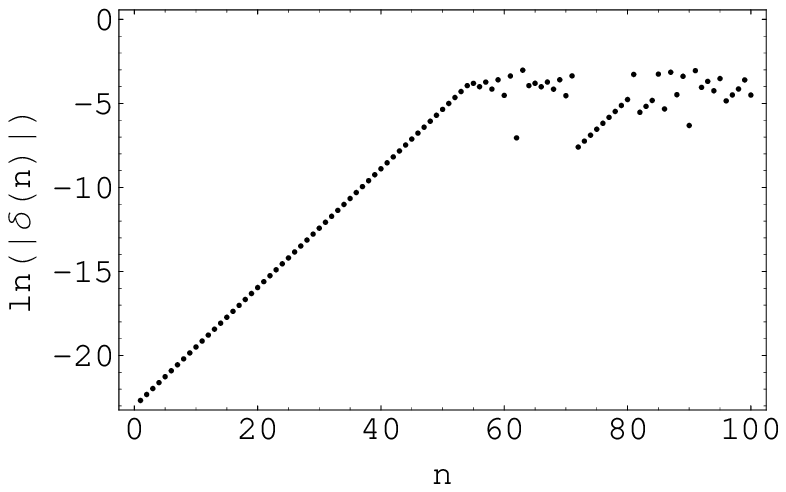}\hspace{1 cm}
\epsfxsize=7cm\epsfbox{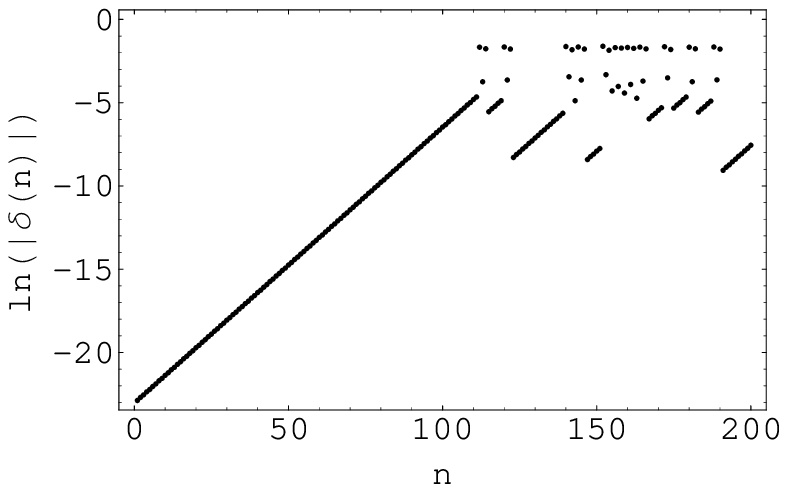}} \caption[ ]{Error growth
and Lyapunov exponent of the application $\f$ given by Eq.
(\ref{e5}) in semi-log plot. Notation here:
$\gaa(n+1)=\f\big[\gaa(n)\big]$. The initial error is
$\delta(0)=10^{-10}$, taken after transients. Left side: $a=0.05$.
From the graph the Lyapunov exponent is $\lambda_{lyap}\simeq
0.353$ while $\ln(\laa)\simeq 0.354$. Right side: $a=0.2$. From
the graph the Lyapunov exponent is $\lambda_{lyap}\simeq 0.166$
while $\ln(\laa)\simeq 0.182$.}
\end{figure}

At this stage we can make the following comments:

\indent\indent\ - The discontinuity of the application $\f$ at $d=0$ is not
the unique origin of its chaotical nature. Indeed, we can
transform $\f$ to be continuous and derivable, e.g. via $a\to a(d)
= a_m\;(1-e^{-\frac{|\gaa|}{\alpha}})$. If $a_m<0.301$ and
$\alpha\ll 1$ then this application exhibits a chaotic behavior.
Let us recall that sudden variation is the source of the chaotic
nature of the application because it permits the folding up of an
expansive application (if $\laa>1$).\\
\indent\indent - When $a>0.311$ the application has stable
fixed points of doubling period. They are obtained from
$\f(\gaa)=-\gaa$.\\
\indent\indent - Increasing $a$, the first separation for $\gaa>0$
inside the interval $\omm$ in two intervals occurs at $a\simeq
0.056$. This result can be retrieved considering the doubling
iterated application $\f^{(2)}=\f\circ \f$, i.e.
$\f^{(2)}(\gaa)=\f\big(\f(\gaa)\big)$. Indeed, this occurs when
$\displaystyle\lim_{\gaa\to 0^-}\f^{(2)}(\gaa)>\gaa^{(2)*}$, i.e.
$\f(a/2)>\gaa^{(2)*}$, where $\gaa^{(2)*}$ is a fixed point of
doubling period where $\gaa^{(2)*}$ is obtained via
$\f(\gaa^{(2)*})=-\gaa^{(2)*}$. From Eq. (\ref{e6}) we retrieve
$a\simeq 1-\frac{2\sqrt{2}}{3} \simeq 0.057$. The extension of
successive iterated points after transients, $\omm$, is then, for
positive values: $]0;\f(a/2)[\;\cup\;
]|\f^{(2)}(a/2)|;a/2[$.\\
\indent\indent - Starting for instance from opinion A being
initially the majority, i.e $\gaa>0$, we evaluate $\gaa_{ch}$,
such as $\f(\gaa_{ch})$=0. Thus, if $\gaa<\gaa_{ch}$ then
$\f(\gaa)<0$ and the majority side will change to the opposite
side; and reciprocally if $\gaa>\gaa_{ch}$ then $\f(\gaa)>0$ and
the majority side will keep the same side (see Fig. 2). From Eq.
(\ref{e6}): \be\label{ed1} \gaa_{ch}\simeq \frac{a}{2\laa}.\ee
Since successive iterated points are contained into the interval
$\omm =]-a/2;a/2[$ and $\f(a/2)<\gaa_{ch}$ with $\laa\leq 1.5$, we
deduce that $\gaa>0$ cannot remain positive more than twice before
turning negative. Note that at second order on $\gaa$,
$\f(a/2)>\gaa_{ch}$ yields a second order equation whose solution
is $\laa>\frac{1+\sqrt{5}}{2}\simeq 1.62$, the golden number.\\


\begin{figure}
\centerline{\epsfbox{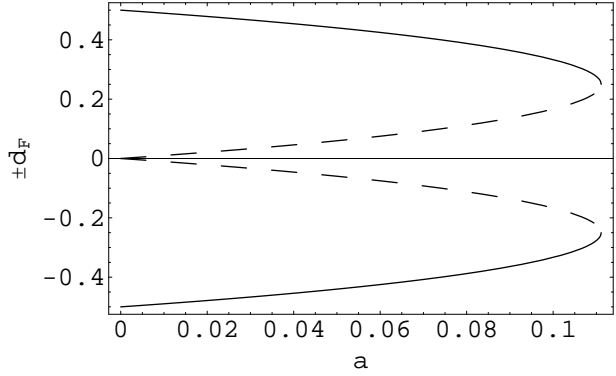}} \caption[ ]{Separators
$\pm \gaa_{F-}$ of chaotic basin in dashed line and point
attractors $\pm \gaa_{F+}$ in plain line, given by Eq
(\ref{fixes}) as functions of density $a$. $a\leq
a_c=\frac{1}{9}$. }
\end{figure}

To summarize, for $a<a_c=\frac{1}{9}$ this model
provides the coexistence of three radically different basins. A
chaotic one located around $\gaa=0$ delimited by separators $\pm
\gaa_{F-}$ and two others at the extremes with each one a point
attractor $\pm \gaa_{F+}$. Fig. (5) shows positions of separators
and points attractors as functions of $a$. For $1/9<a\lesssim
0.301$ only the chaotic basin survives. For $a\gtrsim 0.301$ the opinion forming staggers towards two stable fixed points of
doubling period. Most interesting effects are obtained for a
density of one-sided contrarians such that $a< a_c=\frac{1}{9}$.
Indeed, in this case the initial value $d(0)$ at the beginning of
the electoral campaign determines which one basin is selected. If
$|\gaa(0)|<\gaa_{F-}$ the intention vote dynamics is chaotic but
with a result around $d=0$ such that $d\in ]-a/2;a/2[$. If
$|\gaa(0)|>\gaa_{F-}$, the issue is certain but with a result at
the extremes. This could account for contradictory electoral outcomes
whose results are either around $p=50\%$ or with a huge majority at $p\simeq 80\%$ like the 2002 French presidential election.


\section{Varying the time scales of collective information and local updates}

After having considered 2-step processes, we now study the opinion
dynamics driven by a k-step process where the individual
activation of the one-sided contrarian step occurs after $k-1$
repeated steps of local majority rule update. It accounts for the
fact that polls are not made public every single day during a
campaign while on the contrary people keep on discussing all the
time.

Now $\gaa_{t+k}=D_{o-s}(\gaa_{t+k-1})$ with
$\gaa_{t+k-1}=D_m^{(k-1)}(\gaa_t)$ where $D_m^{(k-1)}=D_m\circ
D_m\circ\cdot\cdot\cdot\circ D_m$, i.e. $k-1$ iterations of $D_m$.
In previous sections, $k$ was equal to 2. Accordingly one-sided
contrarians consider the collective information with a delay
acting at time $t+k-1$ while considering information at time $t$.
However they could as well consider the collective information
without delay, just after the last update inside groups at time
$t+k-1$. Indeed, they act according to $sign[\gaa]$ (see Section
4) and $sign[\gaa_{t}]=sign[\gaa_{t+1}]=\cdots =
sign[\gaa_{t+k-1}]$.

Moreover $\gaa_{t+k-1}$ increases very quickly from the origin
$\gaa=0$, making a contrarian behavior without much effect. For
instance $k=10$ makes the slope at the origin at
$(\frac{3}{2})^{9}\simeq 38$. So, we have to slow down the
dynamics driven by the update rule inside groups in order to study
the varying time scale of collective information. It is done quite
naturally assuming that not every agent eventually changes its
opinion to follow the local majority within each cycle of local
updates turning Eq. (\ref {e1}) to: \be
p_{t+1}=P_{m,w}(p_{t})\equiv w[3p_{t}^2-2p_{t}^3]+(1-w)p_{t},\ee
where $w$ denotes the propensity of an agent to be convinced by
majority rule with $0\leq w\leq 1$. Using the $\gaa$ variable
gives:
\be\label{eDmw}\gaa_{t+1}=D_{m,w}(\gaa_{t})\equiv(1+w/2)\gaa_{t}-2w\;\gaa_{t}^3.\ee

Now $k-1$ iterations of $D_{m,w}$ yields a slower slope at the
origin, e.g. for $w=0.1$ and $k=10$ it is $(1+w/2)^9\simeq 1.5$.

Let $a$ the density of contrarians and $w$ the propensity of
an agent to be convinced by majority rule. The new intention vote
dynamics of the k-step process, generated by the one-sided
contrarian step occurred after $k-1$ repeated local majority rules writes as \be\label{edk}
\gaa_{t+k}=\fn(\gaa_{t})=D_{o-s}\big[D_{m,w}^{(k-1)}(\gaa_{t})\big],\ee
where $D_{o-s}$ and $D_{m,w}$ are respectively given by Eqs.
(\ref{eDo-s}, \ref{eDmw}). This yields at second order on $\gaa_t$
the Eq. (\ref{e6}), but now with the slope
$\laa=(1-a)(1+\frac{w}{2})^{k-1}.$

To exhibit a chaotic behavior the condition for expansion $\laa>1$
gives now: \be k-1>\frac{-\ln(1-a)}{\ln(1+w/2)}.\ee If $a,\; w \ll
1$ and are of the same order, then $k-1>\frac{2a}{w}$, e.g. for
$a=w\ll 1$, $k>3$. Numerically this can be satisfied until
$a=w\leq 0.4$ (see Fig. 6).

With respect of the perpetuity of the chaotic basin, the unstable
fixed points (Eq. (\ref{e7})) are $\pm\gaa_{F-}\simeq
\pm\frac{a}{2(\laa-1)}$, if they exist. It implies $\laa>1$ and
$\laa\leq 1+a$ to have $\gaa_{F-}\leq 1/2$, so,
$k-1>\frac{\ln(\frac{1+a}{1-a})}{\ln(1+w/2)}$. If $a,\; w \ll 1$
and are on the same order, then $k-1>\frac{4a}{w}$, e.g. for
$a=w\ll 1$, $k>5$. Nevertheless, the second order approximation is
not proper as soon as the inequality $\gaa_{F-}\leq 1/2$ doesn't
satisfy $|\gaa|\ll 1$.

To check if successive iterated points cannot escape the chaotic
basin, i.e. if $\gaa_{F-}>a/2$, from Eq. (\ref{e7}) we need to
have $\laa<2$, i.e. \be k-1<\frac{\ln(2)-\ln(1-a)}{\ln(1+w/2)}.\ee
If $a,\; b \ll 1$ and are on the same order, then
$k-1<\frac{2\ln(2)+2a}{w}$, e.g. for $a=w=0.1$, $k<16.9$ while
numerically, $k<18$ (see Fig. 6). As in Section 4, initial value
has to satisfy $|\gaa(t=0)|<\gaa_{F-}$. However, contrarily to the
precedent section, here successive iterated points can escape the
previous chaotic basin.


\begin{figure}
\centerline{ \epsfxsize=7cm\epsfbox{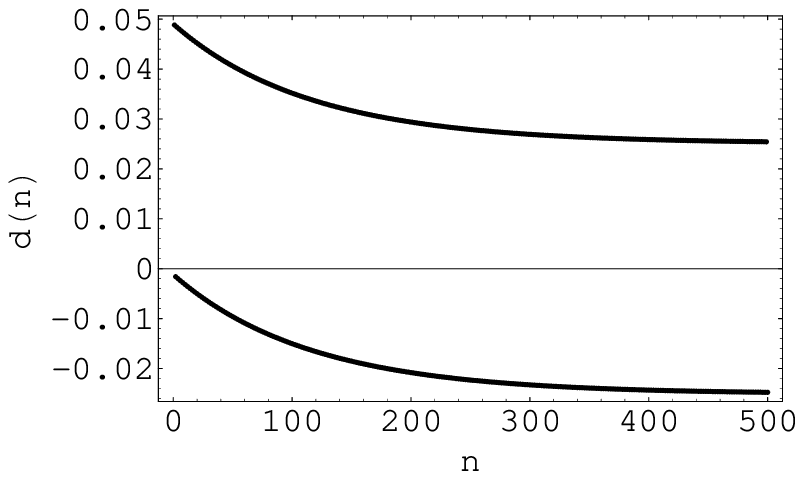}\hspace{1 cm}
\epsfxsize=7cm\epsfbox{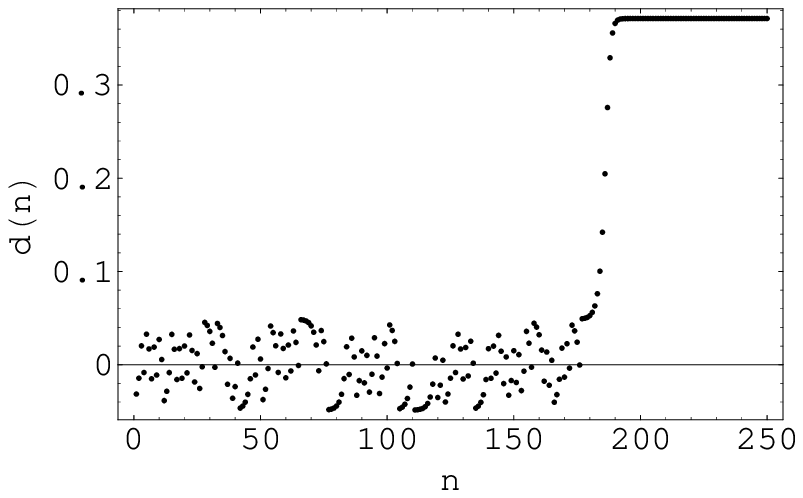}} \caption[ ]{Successive iterated points by the application $\fn$ given by Eq.
(\ref{edk}). Notation here: $\gaa(n+1)=\fn\big[\gaa(n)\big]$. Left
side: $a=w=0.1$, $k=3$ and an initial value $\gaa(0)=0.1$. The
application is not chaotic. $\laa\simeq 0.99<1$. Right side:
$a=w=0.1$, $k=18$ and an initial value $\gaa(0)=-0.04$. The
application is not chaotic by escaping from the previous chaotic
basin. $\laa\simeq 2.06>2$.}
\end{figure}

In addition, for $k>2$, this k-process with $k-1$ repeated
steps of local majority rule updates increase naturally the
majority side persistency before changing compared to the previous
model.


\section{Status quo driven bias}

Up to now both opinions were perfectly symmetric. However while
dealing with political opinion dynamics in view of an election a
difference should be made between the opinion supporting the
current political party in power and the one supporting the
challenging party. The contrarian behavior should be a little more
active against the former winner opinion thus creating a bias in
favor of the challenging opinion \cite{mino}.

Assuming B is the former opinion winner and denoting $\dd$ the
bias in favor of the challenging opinion, Eqs. (\ref{e5},
\ref{e6}) should be rewritten as: \be\label{ebiais}
\gaa_{t+2}=\f(\gaa_t)=(1-a)(\frac{3}{2}\;\gaa_t-2\gaa_t^3)-\frac{a}{2}\;sign[\gaa_t-\dd],\ee
and $\f(\gaa_t)\simeq\lambda_m
\gaa_t-\frac{a}{2}\;sign[\gaa_t-\dd]$ where
$\laa=\frac{3}{2}\;(1-a)$.

In the case $a\leq a_c=1/9$ for which fixed points exist, Fig. (7)
($a=0.05$ and $\dd=0.02$) shows that for $|\dd|<\gaa_{F-}$, the
bias doesn't modify the position of the fixed points. But now,
depending on the ration $\dd$ to $a$, the successive iterated
points could escape from the previous chaotic basin, and thus
reach a point attractor. The values $\dd$, at a fixed density $a$,
for which this new phenomenon occurs, is obtained when
$Sup\big(\displaystyle\lim_{\gaa_t\to
\dd^-}\f(\gaa_t)\;;\displaystyle\lim_{\gaa_t\to\dd^+}|\f(\gaa_t)|\big)>\gaa_{F-}$.

Thus, at first order on $\dd$, $(a/2+\laa |\dd|)>\gaa_{F-}$ and at
second order on $\gaa_t$, Eq. (\ref{e7})
yields: \be\label{esort} |\dd|>\dd_c\simeq a\;\frac{2-\laa}{2\laa
(\laa -1)},\ee where here $1<\laa<\frac{3}{2}$. For instance
$a=0.05$ gives
$\dd_c\simeq 2.37\%$ for the exact value $\dd_c\simeq 2.44\%$.

Evaluating the values $\dd$, at a fixed density $a$, for which
successive iterated points $\gaa_{t}$ have the same sign we find
$|\dd|>\gaa_{ch}$, where $\gaa_{ch}$ is defined as
$\f(\gaa_{ch})=0$ for the application without bias (see Fig. 7).
From Eq. (\ref{ed1}): \be\label{epers}
|\dd|>a\;\frac{1}{2\laa}.\ee For instance $a=0.05$ yields
$|\dd|>1.75\%$ for the exact value $|\dd|>1.76\%$. It is worth
noting that for $\dd$ sufficiently large the application is no
more chaotic. For instance, with $a=\dd=0.2$, the application has
a periodic attractor of period 13.


\begin{figure}
\centerline{ \epsfxsize=7cm\epsfbox{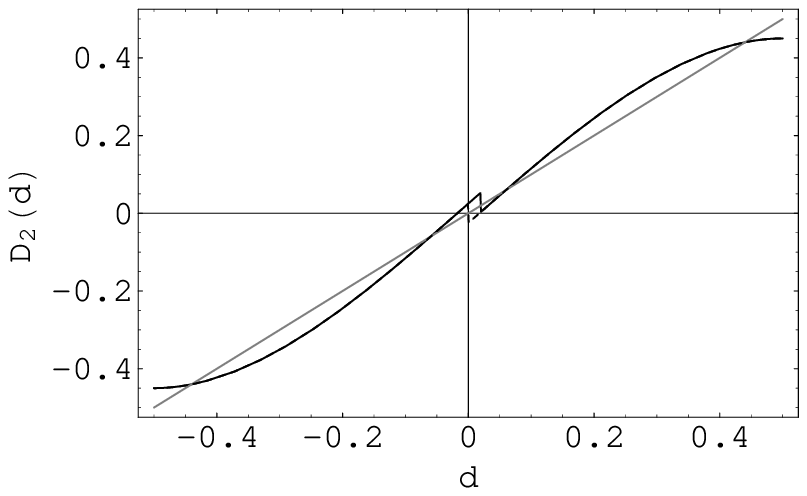}\hspace{1 cm}
\epsfxsize=7cm\epsfbox{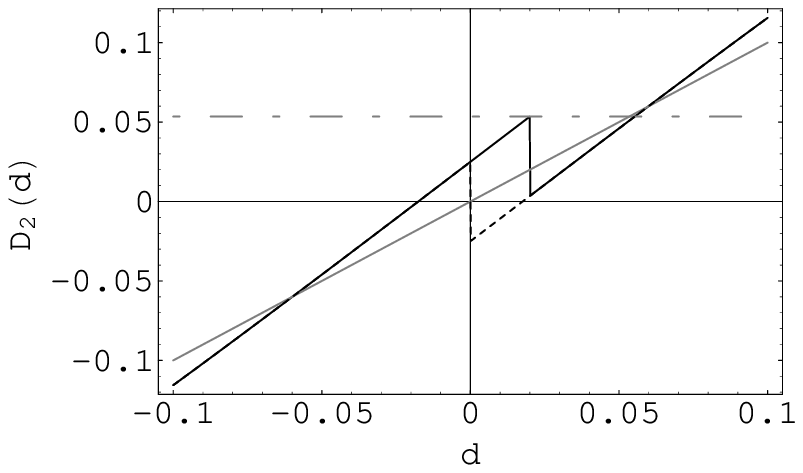}} \caption[ ]{Application
$\f$ given by Eq. (\ref{ebiais}) with a status quo driven bias for
$a=0.05$ and $\dd=0.02$ in plain line. The dashed line is the
application without bias. At the right side, a zoom of this
application around $\gaa=0$. }
\end{figure}

Including a bias in favor of one opinion provides two main
effects. First one, after transients, a majority side (in the
favored opinion) before changing more persistent than without
bias. Second one, for given density $a$ of one-sided contrarians,
it is possible to escape the chaotic basin without bias to reach
the point attractor of the favored opinion.


\section{Unsettled people and contrarian attraction}

We now go back to a population where agents sharing either one
opinion evolve by local majority rule updates only without
contrarians \cite{mino}. But we also consider another population
of agents who do not take part in the public debate. They are
unsettled and hold no opinion. However they are gradually driven
in the public debate on a contrarian basis. At a constant rate $u$
with $0\leq u\leq 1$ they move to the opinion holding population
starting with an opinion opposite to the current majority. Once
they adopt an opinion they become identical to other sharing
opinion agents, i.e., they evolve by local majority rules.

At time $t$ the number of persons sharing opinion A, opinion B and
no opinion are denoted respectively $N_A(t)$, $N_B(t)$ and
$N_{U}(t)$ with $N_O(t)=N_A(t)+N_B(t)$ and $N_O(t)+N_{U}(t)=N$
where $N$ is the total number of agents of both populations.
Associated probabilities are: \be p_{t}=\frac{N_A(t)}{N_O(t)}\quad
and \quad \frac{N_B(t)}{N_O(t)}=1-p_{t}.\ee

We still have a two-step process. The first one is unchanged
with $p_{t+1}=P_m(p_{t})$ while the second one is produced by the
contrarian unsettled agents partial joining the debate. Note that
now the application $p_{t}\rightarrow p_{t+2}$ is no longer
stationary. To account for the shrinking dynamics of unsettled
agents we note $n$ the time which corresponds to the $n^{th}$
iteration i.e. $d_{t_0+2n}\equiv d(n)$ where $t_0$ is the time at
the beginning of the campaign. Thus, writing $d(n+1)=\fnn\big[
\gaa (n) \big]$, it gives: \be\label{eind} d(n+1)=\fnn\big[\gaa
(n)\big]=\frac{\frac{3}{2}d(n)-2d(n)^3}{1+a(n)}-\frac{a(n)}{2\big(1+a(n)\big)}\;sign[d(n)],\ee
with \be\label{e10}
a(n)=u\frac{N_{U}(n)}{N_O(n)}=u\frac{(1-u)^n}{R-(1-u)^n},\ee where
$R=\frac{N}{N_{U}(0)}\geq 1$. For $u\ll 1$, on the first order on
$u\frac{N_{U}(n)}{N_O(n)}$ Eqs. (\ref{e5}, \ref{e6}) are unchanged
with now instead of $a$ an effective time dependent contrarian
density: $a\rightarrow a(n)$.


\begin{figure}
\centerline{\epsfbox{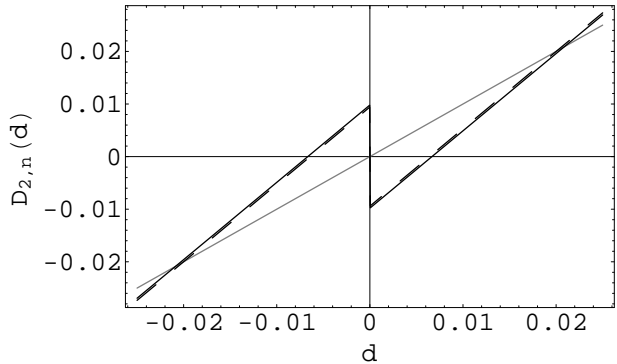}} \caption[ ]{Applications $\fnn$
given by Eq. (\ref{eind}) with $u=0.02$ and $R=2$. $n=0$ in plain
line and $n=1$ in dashed line.}
\end{figure}

In the limit $u\ll 1$ from Eq. (\ref{e7}) we can write,
$\gan\simeq a(n)$ with $\omn$ defined for
$0<|\gaa|<\frac{a(n)}{2}$. With $\displaystyle\lim_{n\to
\infty}a(n)=0$ the interval $\omn$ shrinks to $0$. However this
model prohibits $n\to\infty$ since the number of unsettled persons
mobilized at each iteration is an integer and not a real one,
although the $n^{th}$ iteration, the number $uN_{U}(0)(1-u)^n$ is
assumed to be greater than $1$.

In the case $(R-1)\gg u$, $ a(n+1)\simeq
\big(1-u\;\frac{R}{R-1}\big)\;a(n)$ from Eq. (\ref{e10}).
Accordingly points escape the basin bordered by $\pm \gaa_{n,F-}$
from time $n$ to $n+1$, if $|\gaa (n+1)|>\gaa_{n+1,F-}\simeq
a(n+1)$ implying $|\gaa(n)|>(1-u\;\frac{3R}{2(R-1)})a(n)$ (see
Fig. 8). Thus, if $d$ belongs to the interval $\omn$ at time $n$,
successive iterated points will be contained into the successive
intervals $\omn$. (see Fig. 9).

Even if successive iterated points do not escape the basin
bordered by $\pm \gan$, the dynamics is no longer chaotic. Indeed
the point $\gaa=0$ is now asymptotically stable. This is due to
the non-stationary dynamics effects. Nevertheless, the shrinking
dynamics doesn't affect the sign of iterated $\gaa$. So, the
expected winner at the issue of an electoral campaign remains
unpredictable.

Including unsettled people driven gradually in the public
debate on a contrarian basis provides a shrinking of the chaotic
basin of the one-sided contrarians model. Thus, the results may
shed light to recent very unusual elections like the hung 2000 USA
presidential and German 2005 elections. Indeed, the winners of
these long electoral campaigns were very unpredictable and the
outcomes very tied.


\begin{figure}
\centerline{\epsfbox{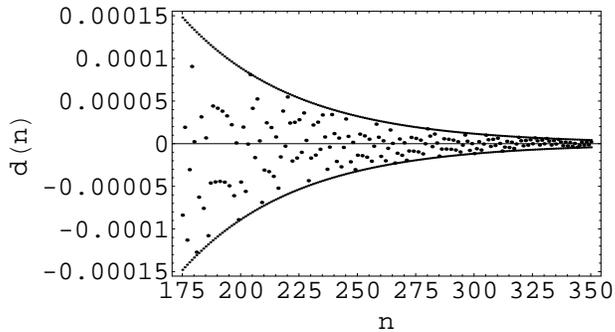}} \caption[ ]{Successive iterated points with $\fnn$ given by Eq. (\ref{eind})
where $u=0.02$ and $R=2$. Borders $\pm \frac{a(n)}{2}$ are
included in thin points. Notation here: $d(n+1)=\fnn\big[ \gaa (n)
\big]$.}
\end{figure}

\section{Conclusion}

We have presented a simple model giving a deterministic opinion
dynamics (see Eq. \ref{e5}), which can be chaotic. Furthermore,
chaotic basin can coexist with two point attractors at the
extremes. This model contains two main effects. First effect,
amplification one given by the local majority rule inside groups.
Second effect, retroaction one given by the action of the
one-sided contrarians. The one-sided contrarians act by comparison
and opposition to a collective information, the majority. Their
action introduces to dynamics a discontinuity.

Afterwards, rooted in this simple model, some others features of
electoral campaigns are added, like the fact that polls are not
made public every day, or a bias in favor of one opinion, or the
influence of unsettled people gradually driven in the public
debate on a contrarian basis.

To sum up our various results we come back to the one-sided
contrarian model deterministic equation used throughout this
paper: \be
f(x)=(1-a)(\frac{3}{2}\;x-2x^3)-\frac{a}{2}\;sign[x],\ee where
$x\equiv d\in[-1/2;1/2]$ and $a$ is the density of one-sided
contrarians, $0\leq a\leq 1$. At the second order on $x$
approximation, with $\laa=\frac{3}{2}(1-a)$, $f(x)\simeq
\laa\;x-\frac{a}{2}\;sign[x]$. $\laa$ is the expansion effect. The
term $-\frac{a}{2}\;sign[x]$ is related to the folding up, i.e.
the retroaction effect. The term $-x^3$ can be seen as a
non-linear saturation effect and gives the point attractors. It is
symmetric with respect to $x=0$.

Afterwards,
accounting for some others social parameters enrich the model and modify a little bit its
applications. By decoupling the local update time sequence from the
one-sided contrarian behavior activation, the two effects,
amplification and folding up effects, can act separately. Next, a
bias in favor of one opinion introduced as a simple parameter
$\dd$ fixes the discontinuity position at $x=\dd$, i.e.,
$sign[x]\to sign[x -\dd]$. The opinion dynamics is no more
symmetric. Last, unsettled people driven gradually to public
debate on a contrarian basis roughly plays with the folding up
effect uniquely; $\laa \simeq \frac{3}{2}$ and $a\to a(n)$ with
$a(n)\to 0$ when $n\to \infty$. The opinion forming dynamics is no
more stationary.

At this stage it is worth to stress that contrary to what could be expected, our treatment using probabilities, local updates and reshuffling between updates does not define a mean field like frame. This result was demonstrate recently using Monte Carlo simulations of a nearest neighbor ferromagnetic Ising system on a square lattice \cite{new}. Indeed it creates a new class of universality in addition to both 2-d Ising and mean field ones. For our current model, in the case of group size four with no contrarians, using a cellular automata was shown to recover our analytical result \cite{chopard}.

To conclude it is worth to notice that in terms of real life situations, due to the
finite number of iterations imposed by the fact that any public
campaign is finite in time, the intention vote dynamics will not
exhibit a chaotic behavior although in principle it could. In
addition, the non zero fuzziness of poll measurements of the
initial intention vote distribution, result automatically in a
growing error making difficult to predict the sign of successive
iterations (see Fig. 10).

Although this simple model does not pretend to account
for an exhaustive explanation of opinion forming during electoral
campaigns it exhibits some features which could shed new light on
recent surprising voting outcomes, like for instance, on the one
hand, an unpredictable issue with a very tied outcome
like the German 2005 and the USA 2000 elections, and on the other
hand, a well predicted outcome with a huge majority like in the
2002 French presidential elections with a majority
around $80\%$.

However it is worth to stress that at this stage if we are able to embody above contradictory voting outcomes within a single frame, we are not in a position to select the prevailing one. To overpass this difficulty would require the collaboration with social scientists to estimate the actual type of contrarians involved in a given election. It is open for future work.


\begin{figure}
\centerline{\epsfbox{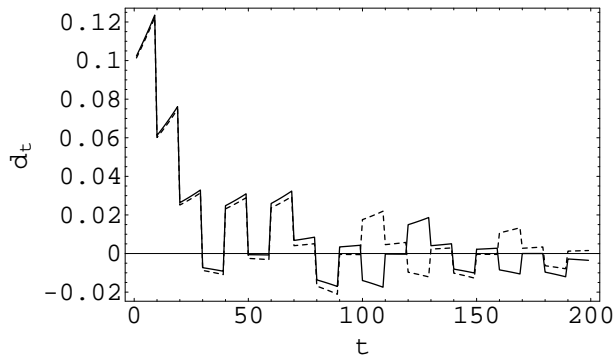}} \caption[ ]{Combination of a
k-step process and unsettled agents with $k=10$, $w=0.05$,
$u=0.05$ and $R=1.5$. Initial values are $0.1$ in plain line and
$(0.1-0.001)$ in dashed line. }
\end{figure}

{\bf Acknowledgments}

The authors are grateful to thank Maurice Courbage for helpful and
fruitful discussions.

\end{document}